\documentclass[12pt]{iopart}
\usepackage{epsfig}

\begin{document}

\title[]{Ellipsometric measurements of the refractive indices of linear alkylbenzene and EJ-301 scintillators from 210 to 1000 nm}

\author{H.~Wan Chan Tseung and N.~Tolich}

\address{Center for Experimental Nuclear Physics and Astrophysics, and Department of Physics, University of Washington, Seattle, WA 98195}
\ead{hwan@uw.edu}

\begin{abstract}

\noindent We report on ellipsometric measurements of the refractive indices of LAB-PPO , Nd-doped LAB-PPO and EJ-301 scintillators to the nearest $\pm 0.005$, in the wavelength range $210$--$1000$ nm.
\end{abstract}

%\noindent{\it Keywords}: refractive index, linear alkylbenzene.
\submitto{\PS}
%\maketitle

\section{Introduction}
There is presently a lot of interest in using linear alkylbenzene (LAB) as a scintillator solvent for neutrino experiments ({\it e.g.} SNO\protect\raisebox{.15ex}{+}, LENA, RENO and Daya Bay \cite{SNO+, LENA, RENOProp, DayaBay}). First proposed as a scintillator  by M. Chen, this organic liquid has the advantages of low toxicity, good light yield, high flash-point and long attenuation length. Furthermore, it is compatible with acrylic, a material that is frequently used to build the scintillator containment vessel. 

To understand the propagation of light and reconstruct events in these experiments, and to accurately calculate the Cherenkov light production, it is imperative to know the refractive index of the liquid scintillator as a function of wavelength. Currently, our knowledge of the refractive index of LAB-based scintillators is limited to the optical region. Recently, the RENO collaboration reported a measurement of the refractive index of LAB at six wavelengths in the interval $400<\lambda<630$ nm, using the mininum deviation technique \cite{Reno}. In this paper, we describe a measurement of the real component of the refractive index, $n$, of two LAB-based scintillators, to the nearest $\pm 0.005$ at 500 points from 210 to 1000 nm, using an ellipsometric method \cite{Synowicki, Costner}. These two scintillators were: LAB with 3 g/L of 2,5-diphenyloxazole (PPO), and LAB-PPO loaded with 0.094 \% Nd by weight.

We also report on a measurement of $n$ for EJ-301 scintillator \cite{Eljen}, which is equivalent to the very commonly-used NE-213 and BC-501A scintillators. As far as we know, the dispersion of NE-213 has not been measured down to the UV region.

\section{Experimental method}
The use of ellipsometers for measuring the refractive indices of liquids is convenient, but not widespread. Because of challenges brought about by surface vibrations, evaporation and difficulties in controlling the temperature of the liquid medium, the ellipsometric method is not expected to be as competitive as the minimum deviation technique for precision refractive index measurements. However, measurements that are accurate to the third decimal place are possible. In this work, we adopted the simplest ellipsometric configuration, in which polarized light reflects off an infinitely deep layer of liquid. This is described below.

\subsection{Theory}\label{section:theory}
\begin{figure}[!hpt]
\centering
\includegraphics[width=7.7cm]{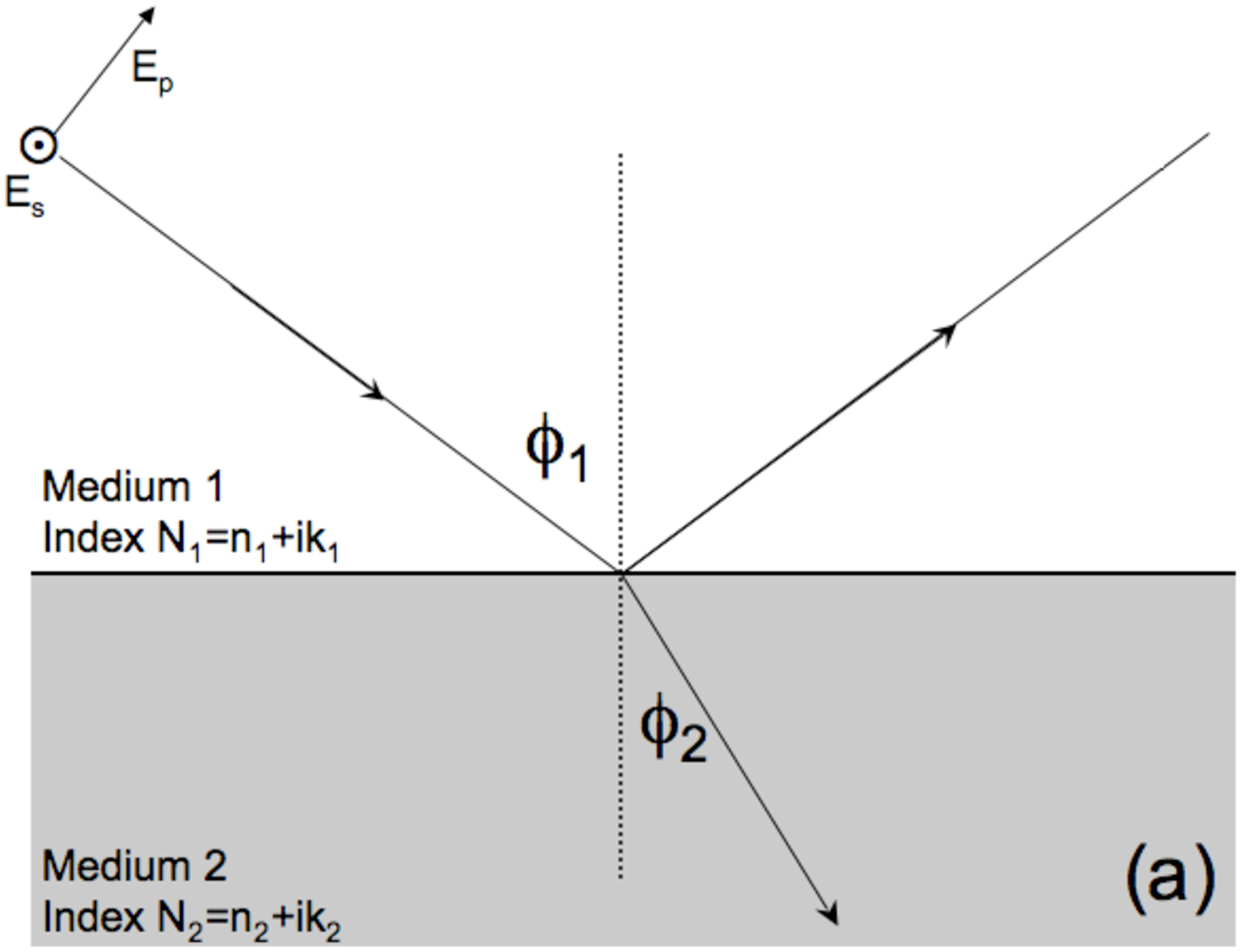}
\includegraphics[width=7.7cm]{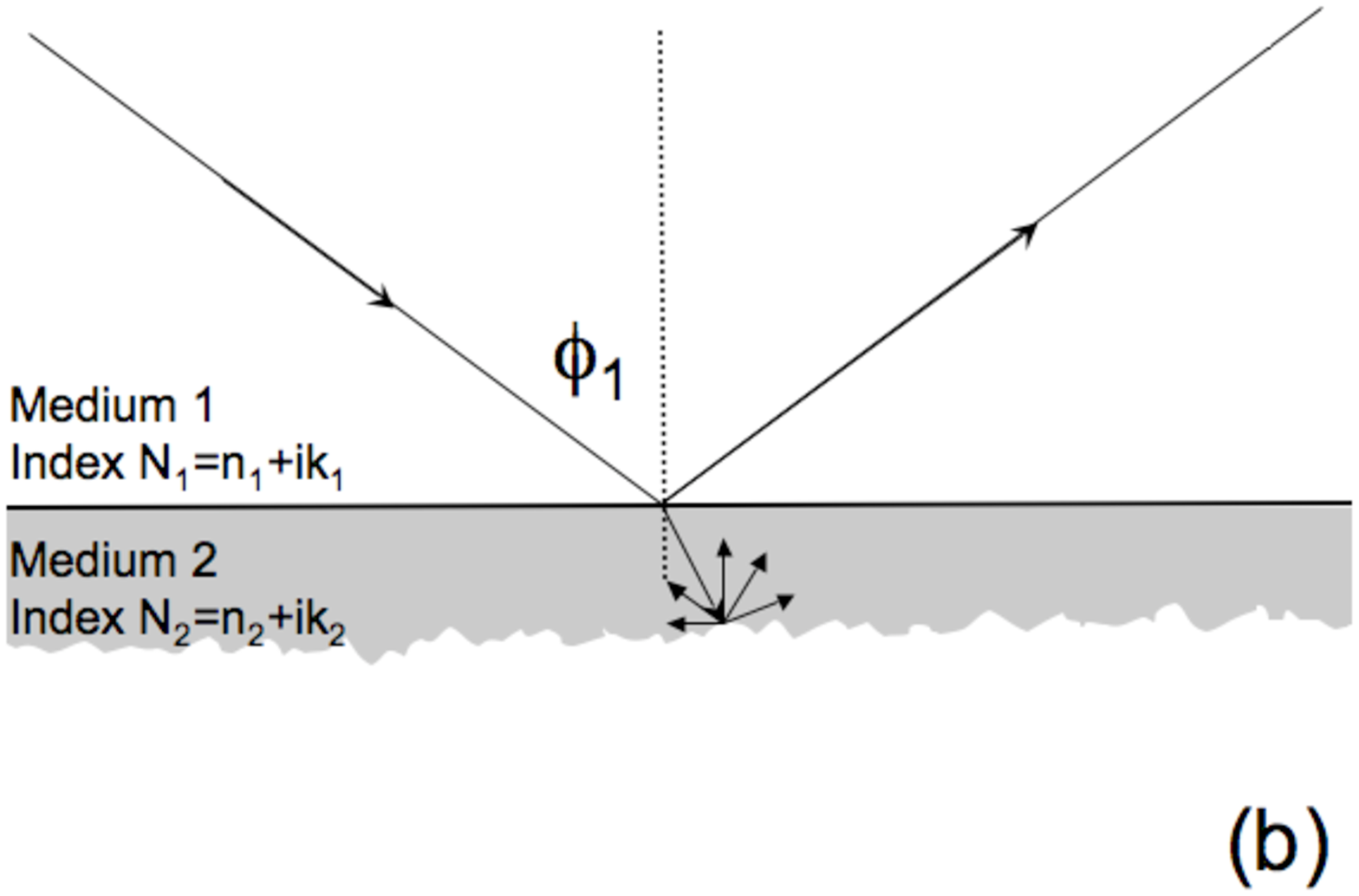}
\caption[]
 {(a) Definition of parameters used in \S\ref{section:theory}. (b) Set-up with liquid film on a frosted glass slide. This is equivalent to a deep container.
}
\label{figure:reflectionfigures}
\end{figure}

Consider polarized light incident on the interface between two media, both of which are infinite and homogeneous (figure \ref{figure:reflectionfigures}(a)). After reflection, there is a change in polarization that is dependent on the incident angle, the refractive indices of the two media, as well as the initial polarization state. The ellipsometer measures the change in polarization in terms of two parameters, $\Delta$ and $\Psi$. If $R_p$ and $R_s$ are the reflection coefficients of the electric field components parallel ($E_p$) and perpendicular ($E_s$) to the plane of incidence, the ellipsometric parameters are given by:
\begin{equation}\label{equation:ellipsometer} \frac{R_p}{R_s}=\tan\Psi\mathrm{e}^{i\Delta}\end{equation}
$R_p$ and $R_s$ are expressed in terms of the complex refractive indices of media 1 and 2 ($N_1$ and $N_2$),  and the angles $\phi_1$ and $\phi_2$ (figure \ref{figure:reflectionfigures}(a)) by Fresnel's equations. Combining Fresnel's equations and Snell's law with equation \ref{equation:ellipsometer}, and separating real and imaginary components, one gets \cite{Tompkins}:
\begin{equation}\label{equation:n21} n_{2}^2-k_2^2=n_1^2\sin^2{\phi_1}\left[1+\frac{\tan^2\phi_1(\cos^2 2\Psi-\sin^2\Delta\sin^2 2\Psi)}{(1+\sin 2\Psi \cos \Delta)^2}\right]=X\end{equation}
\begin{equation}\label{equation:n22} 2n_2k_2=\frac{n_1^2\sin^2\phi_1\tan^2\phi_1\sin 4\Psi\sin\Delta}{(1+\sin 2\Psi\cos \Delta)^2}=Y\end{equation}
where $n_1$ and $n_2$ are respectively, the real components of $N_1$ and $N_2$, and $k_2$ is the imaginary component of $N_2$. Given $\Delta$, $\Psi$, $\phi_1$ and $n_1$, one can therefore solve for $n_2$ and $k_2$ separately:
\begin{equation}\label{equation:n23} n_2^2=\frac{1}{2}(X+\sqrt{X^2+Y^2})\quad,\quad k_2^2=\frac{Y}{2n_2}\end{equation}
In our case, medium 1 is air, and medium 2 is the liquid scintillator, whose real refractive index component ($n_2$) is to be measured. The sensitivity to $k_2$ is weak, since the measurement is made on the reflected, and not the transmitted light component, which conveys most of the information on attenuation in medium 2.

\subsection{Set-up}

For equations \ref{equation:n21} and \ref{equation:n22} to be valid, one must have: (1) an ideal interface, and (2) an infinitely deep medium 2 (so that no interference occurs with light reflected off the bottom of medium 2). To emulate a deep container with a flat liquid surface, we followed Synowicki {\it et al.} \cite{Synowicki}: a few drops of liquid were allowed to spread into a thin film, on a sandblast-frosted microscope glass slide (figure \ref{figure:reflectionfigures} (b)). Measurements were then made with the light reflected specularly off the upper surface of the liquid film. The transmitted component was diffusely reflected off the frosted glass and did not affect the results. The slides, as well as all other glassware used in this work, were first cleaned in an ultrasound bath. Data was taken at five values of $\phi_1$, namely $55^{\circ}, 60^{\circ}, 65^{\circ}, 70^{\circ}$ and $75^{\circ}$.  The experiment was performed at room temperature (25$^{\circ}$ C) with a variable-angle Woollam M2000 ellipsometer, fitted with a Xe lamp. Excluding alignment procedures, a spectral scan at one angle typically took  around 30 s. 
%\subsection{Setup and procedure}

\section{Results}

\subsection{LAB-PPO and Nd-doped LAB-PPO}
\begin{figure}[!hpt]
\centering
\includegraphics[width=7.5cm]{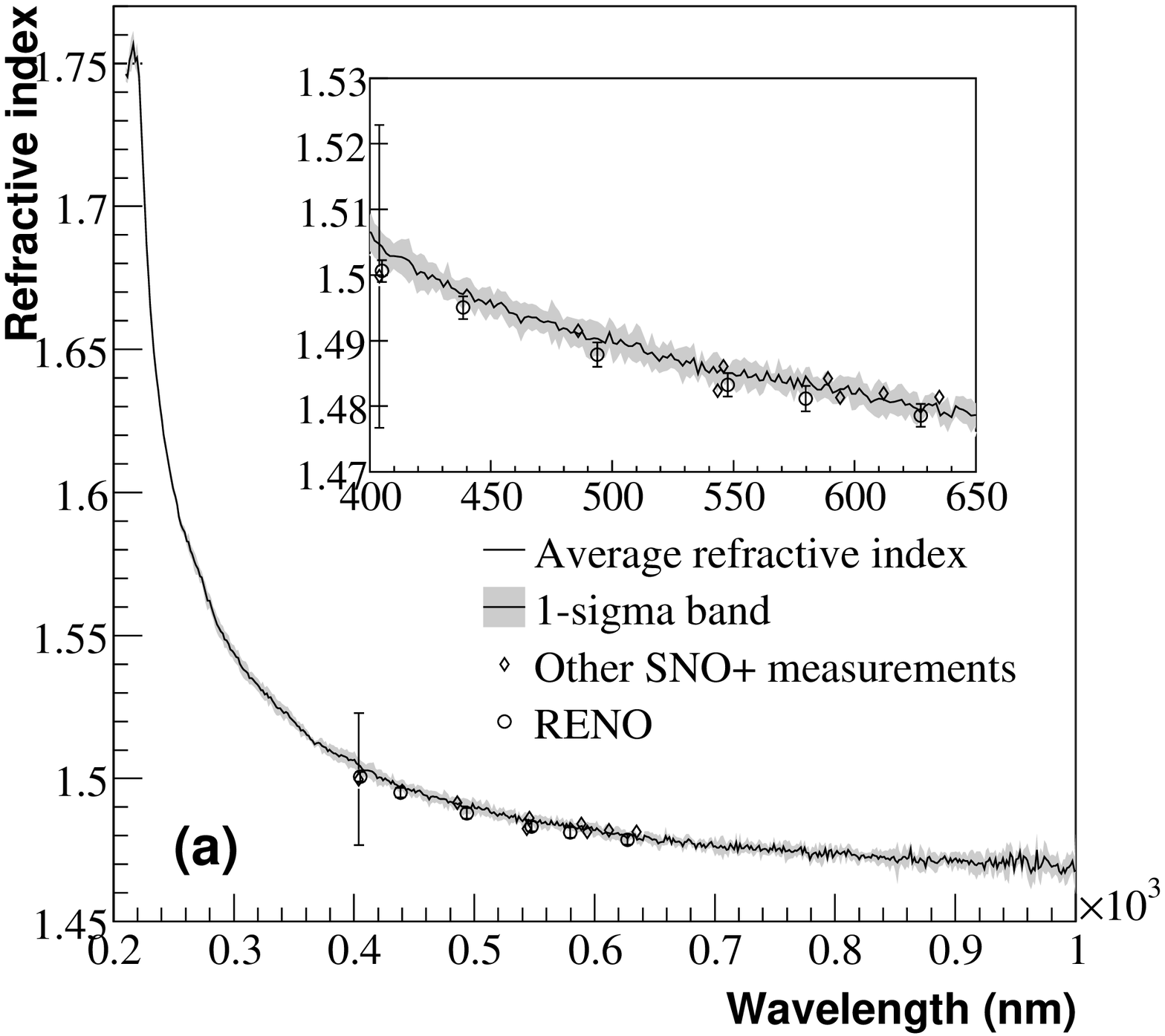}
\includegraphics[width=7.5cm]{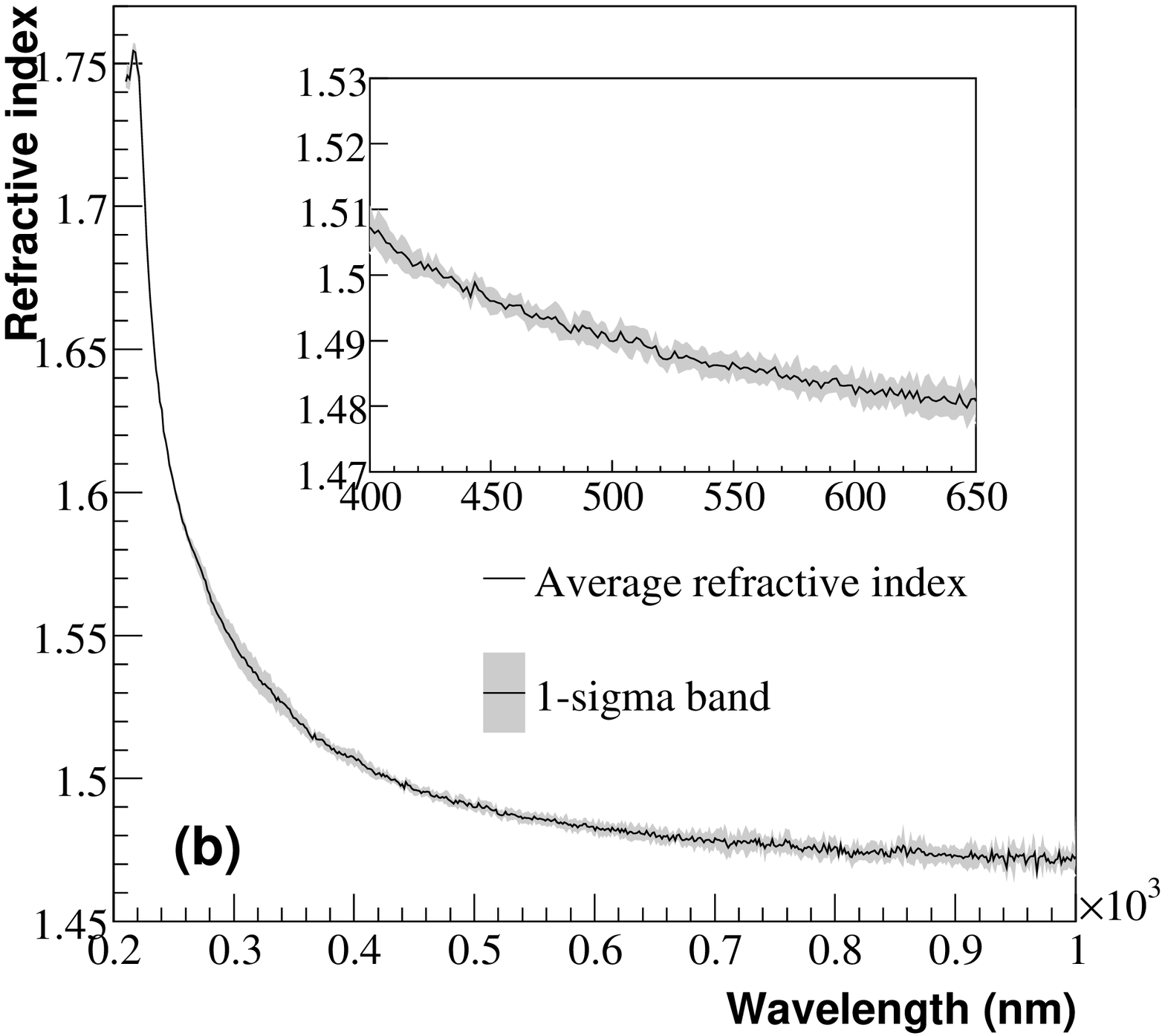}
\caption[]
 {Measured real refractive index components for (a) LAB-PPO, and (b) Nd-doped LAB-PPO. Previous measurements for LAB-PPO in the optical region are also shown for comparison. The insets show the interval  400--650 nm in more detail.
}
\label{figure:LABresults}
\end{figure}

Results for LAB-PPO are shown in figure \ref{figure:LABresults}(a). The measurements at the five different angles agree within around $\pm 0.005$ over 200--1000 nm. The solid curve is the average from the five angles, while the shaded region shows the standard deviation. Shown in the circle markers are measurements from RENO \cite{Reno}, obtained using the mininum deviation technique. The diamond markers are previous measurements by the SNO\protect\raisebox{.15ex}{+}  collaboration \cite{SNOindex}, taken using an Abbe refractometer and other methods. The Nd-doped LAB-PPO results are shown in figure \ref{figure:LABresults}(b). The values are rather close to undoped LAB-PPO, the difference being within $\pm$0.5 \% over the entire wavelength range under study. 

\begin{figure}[!hpt]
\centering
\includegraphics[width=7.5cm]{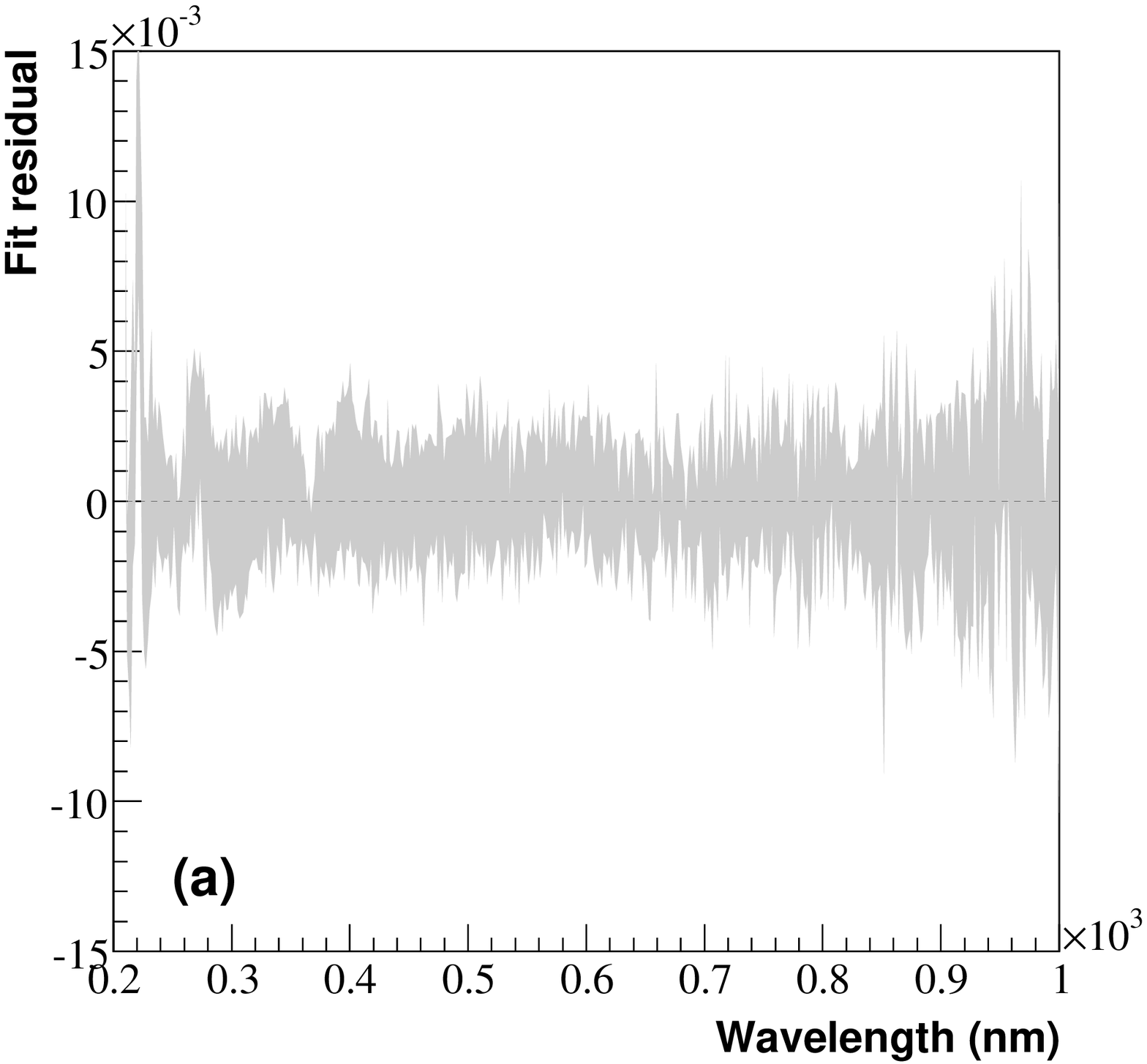}
\includegraphics[width=7.5cm]{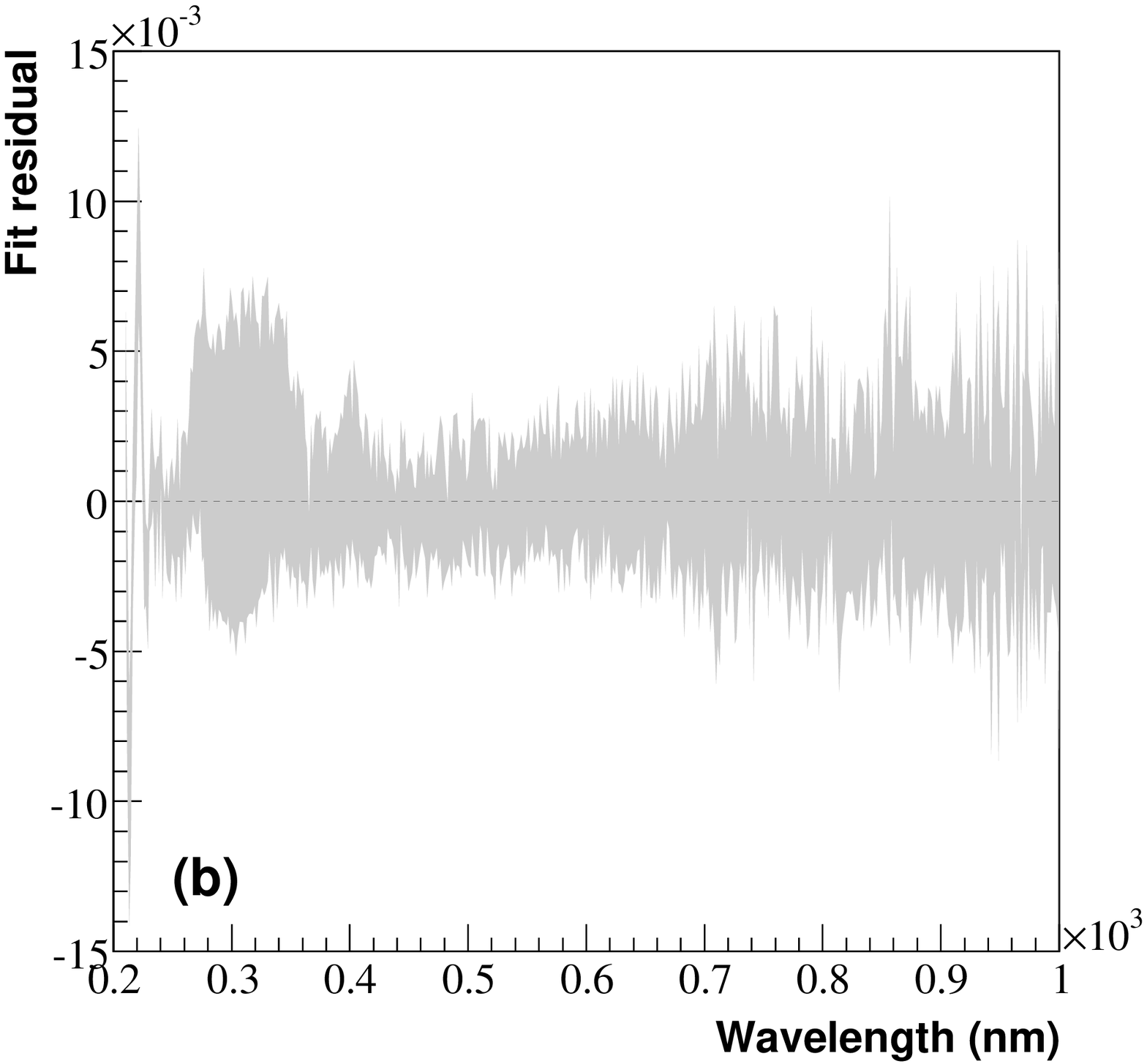}
\caption[]
 {Residuals of the fits to the Sellmeier formula (Eq.~\ref{equation:sellmeier}) in the range 230--1000 nm and to a 3$^{rd}$ order polynomial in the range 210--230 nm. (a): LAB-PPO. (b): Nd-doped LAB-PPO.}
\label{figure:fitresiduals}
\end{figure}

For the wavelength range 230--1000 nm, where the refractive indices of both liquids decrease monotonically with wavelength $\lambda$, we parameterize the dispersion with the Sellmeier formula:
\begin{equation}\label{equation:sellmeier} n^2(\lambda) = 1+\sum_{i=1}^N \frac{B_i}{1-\left(\frac{C_i}{\lambda}\right)^2}\end{equation}
with $N=4$. For the range 210--230 nm, we fit to a third order polynomial:
\begin{equation}\label{equation:polynomial} n = A_0+A_1 \lambda + A_2 \lambda^2 + A_3 \lambda^3 \end{equation}
In both equations \ref{equation:sellmeier} and \ref{equation:polynomial}, $\lambda$ is in nm. The Sellmeier and polynomial coefficients are shown in tables~\ref{table:sellmeiercoeff} and \ref{table:polycoeff}, respectively. The fit residuals are shown in figure \ref{figure:fitresiduals}.
%The data at 75$^{\circ}$ deviates rather markedly from those taken at the other four angles. This could have been due to an unintentional change of alignment ({\it e.g.} sample movement) during the data-taking.

\begin{table}
\caption{\label{table:sellmeiercoeff} Sellmeier coefficients for LAB-PPO, Nd-doped LAB-PPO and EJ-301. The LAB-PPO and Nd-doped LAB-PPO results are valid for the wavelength range 230--1000 nm, while the EJ-301 results apply to the range 300--1000 nm. The $C$ coefficients are given in nm.}
\footnotesize\rm
\begin{tabular*}{\textwidth}{@{}ccccccccc}
\br
Liquid&$B_1$&$C_1$&$B_2$&$C_2$&$B_3$&$C_3$&$B_4$&$C_4$\\
\mr
LAB-PPO&0.821384&94.7625&0.311375&160.751&0.0170099&219.575&0.608268&9385.54\\
Nd LAB-PPO&0.80559&98.7814&0.325456&157.743&0.0175545&219.332&7.53674$\times$$10^{-5}$&7301.7\\
EJ-301&0.715235&1.29988&0.464991&203.739& 0.0302529&0.484641&1.99773&14719.6\\
\br
\end{tabular*}
\end{table}

\begin{table}
\caption{\label{table:polycoeff}  Coefficients for 3$^{rd}$ order polynomial fits to LAB-PPO, and Nd-doped LAB-PPO refractive indices in the wavelength range 210--230 nm.}
\begin{indented}
\footnotesize\rm
\item[]\begin{tabular}{@{}ccccc}
\br
Liquid&$A_0$&$A_1$&$A_2$&$A_3$\\
\mr
LAB-PPO&-209.56&2.81043&-0.0124283&$1.82697\times 10^{-5}$\\
Nd LAB-PPO&-206.13&2.76412&-0.0122203&$1.79591\times 10^{-5}$\\
%EJ-301&\\
\br
\end{tabular}
\end{indented}
\end{table}

\subsection{EJ-301}
\begin{figure}[!hpt]
\centering
\includegraphics[width=7.5cm]{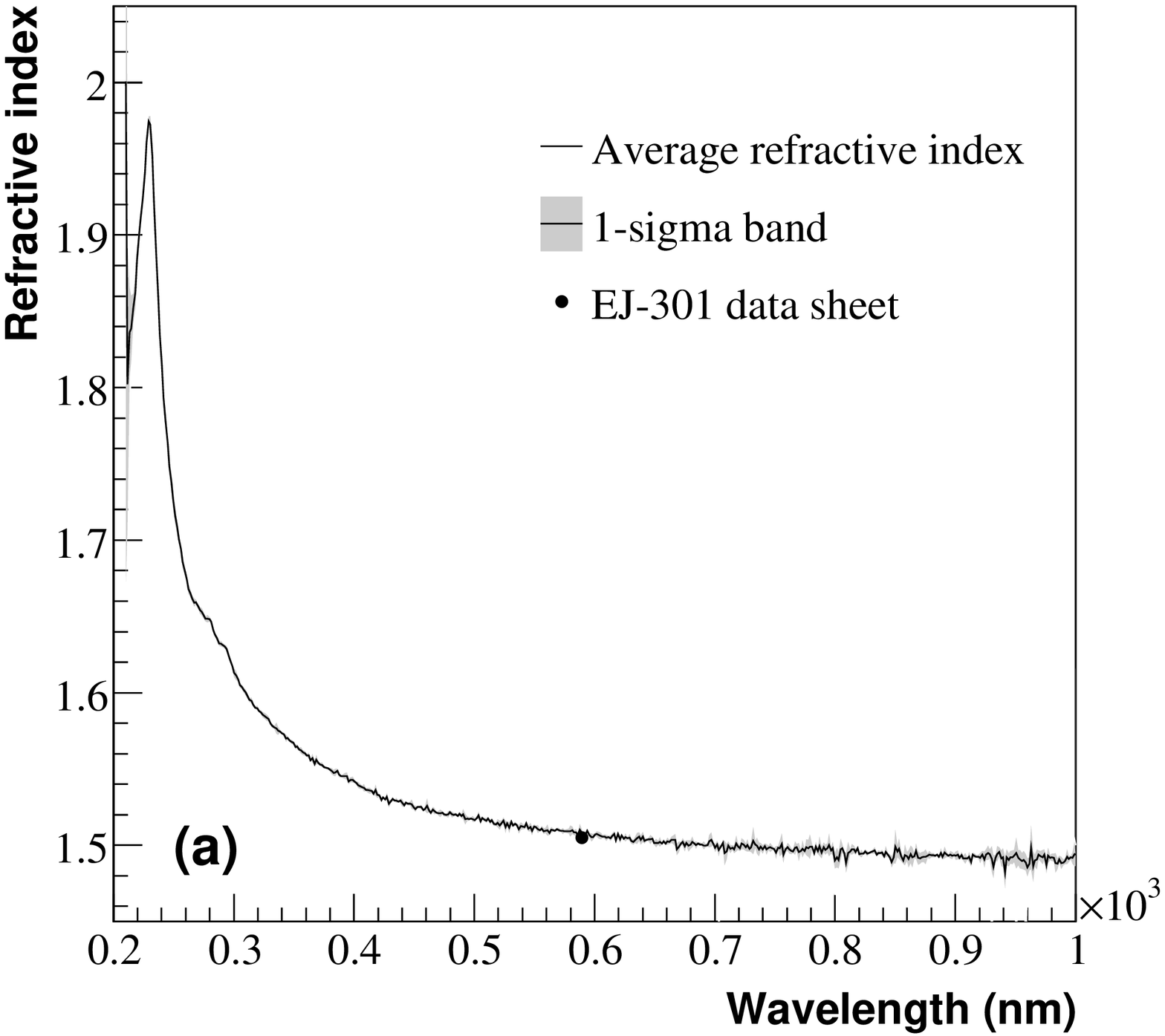}
\includegraphics[width=7.5cm]{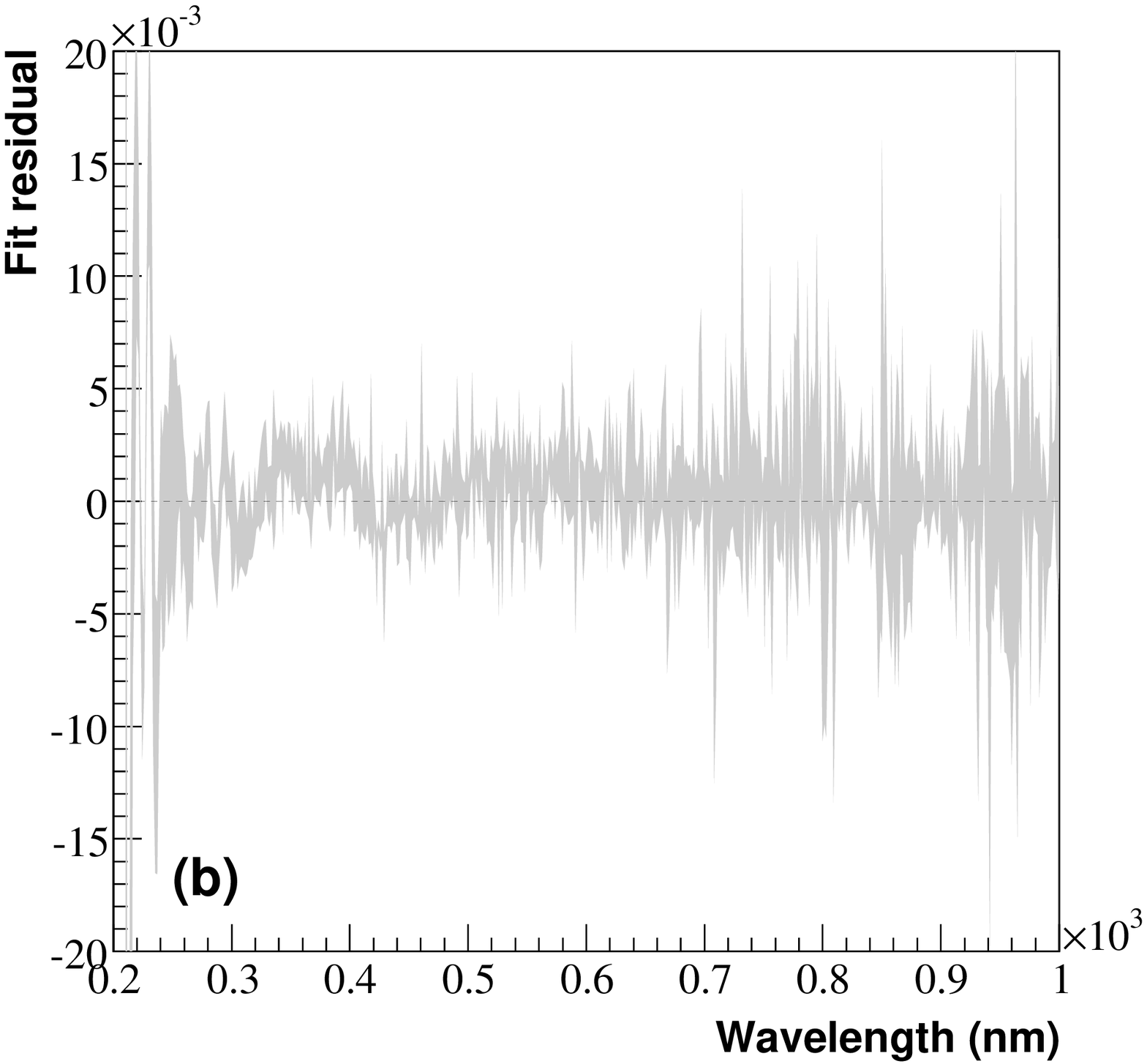}
\caption[]
 {(a): Measured real refractive index components for EJ-301. The value at 589.3 nm is from \cite{Eljen}. (b): residuals of fits to two 4$^{th}$ order polynomials in the wavelength regions 210--240 nm and 240--300 nm, and to the Sellmeier equation in the region 300--1000 nm.}
\label{figure:NE213results}
\end{figure}

The EJ-301 results are shown in figure \ref{figure:NE213results}(a). There appears to be a region of anomalous dispersion just below 230 nm. The measurement uncertainty increased markedly below 220 nm. To parameterize the EJ-301 refractive index, we fit to the Sellmeier equation (equation \ref{equation:sellmeier}, with $N$=4) in the region 300--1000 nm, and to two separate 4$^{th}$ order polynomials in the regions 210--240 nm and 240--300 nm. The Sellmeier coefficients are given in the third row of table \ref{table:sellmeiercoeff}, while the 4$^{th}$ order polynomial coefficients are shown in table \ref{table:ne213polycoeff}. The fit residuals are shown in figure \ref{figure:NE213results}(b).

\begin{table}
\caption{\label{table:ne213polycoeff}  Coefficients for 4$^{th}$ order polynomial fits to the EJ-301 refractive index in the wavelength ranges 210--240 nm and 240--300 nm.}
%\begin{indented}
\footnotesize\rm
\begin{tabular}{@{}cccccc}
\br
Wavelength range&$A_0$&$A_1$&$A_2$&$A_3$&$A_4$\\
\mr
210--240 nm&11784.7388&-205.807243&1.34648820&-3.9107264$\times$10$^{-3}$&4.25436903$\times$10$^{-6}$\\
240--300 nm&183.668&-2.51192&0.0129879&-2.98047$\times$10$^{-5}$&2.55974$\times$10$^{-8}$\\
%EJ-301&\\
\br
\end{tabular}
%\end{indented}
\end{table}

%\subsection{Imaginary refractive index component}
%The ellipsometric method is not sensitive to $k_2$, which is better determined by transmittance measurements. The value of $k_2$, which is close to 0 over a wide wavelength range, is proportional to $\sin\Delta$ (Eq.~\ref{equation:n23}). In the present work, large uncertainties were obtained on $k_2$ due to $\Delta$ being close to either $0^{\circ}$ or $180^{\circ}$ (as expected from a dielectric \cite{Tompkins}). The precision in the determination of $\Delta$ is insufficient to make a definite statement on $k_2$.

\section{Conclusions}
We performed ellipsometric measurements of the refractive indices of EJ-301, LAB-PPO and Nd-doped LAB-PPO scintillators between 210 and 1000 nm. Our measurements are accurate to $\pm 0.005$. Empirical formulae for the index of each scintillator are given.

\section{Acknowledgements}
This work was supported by the United States Department of Energy grant \#DE-FG02-97ER41020. We thank our Brookhaven National Laboratory (BNL) collaborators for sending us the LAB scintillator samples, and S. Asahi and M. Yeh for sharing the measurements performed at Queen's University and BNL in the optical region. 
Part of this work was conducted at the University of Washington NanoTech User Facility, a member of the NSF National Nanotechnology Infrastructure Network (NNIN).

\end{document}